\documentclass[sigconf,screen]{acmart}

\AtBeginDocument{%
  \providecommand\BibTeX{{%
    \normalfont B\kern-0.5em{\scshape i\kern-0.25em b}\kern-0.8em\TeX}}}

\copyrightyear{2023}
\acmYear{2023}
\setcopyright{rightsretained}
\acmConference[UIST '23]{The 36th Annual ACM Symposium on User Interface Software and Technology}{October 29-November 1, 2023}{San Francisco, CA, USA}
\acmBooktitle{The 36th Annual ACM Symposium on User Interface Software and Technology (UIST '23), October 29-November 1, 2023, San Francisco, CA, USA}
\acmDOI{10.1145/3586183.3606773}
\acmISBN{979-8-4007-0132-0/23/10}

\begin{document}

\title[CrossTalk: Intelligent Communication and Collaboration Media]{CrossTalk: Intelligent Substrates for Language-Oriented Interaction in Video-Based Communication and Collaboration}

\author{Haijun Xia}
\email{haijunxia@ucsd.edu}
\affiliation{%
  \institution{University of California San Diego}
  \streetaddress{9500 Gilman Dr}
  \city{La Jolla}
  \state{California}
  \country{USA}
  \postcode{92093}
}

\author{Tony Wang}
\email{yiw136@ucsd.edu}
\affiliation{%
  \institution{University of California San Diego}
  \streetaddress{9500 Gilman Dr}
  \city{La Jolla}
  \state{California}
  \country{USA}
  \postcode{92093}
}

\author{Aditya Gunturu}
\email{adigunturu@outlook.com}
\affiliation{%
  \institution{University of California San Diego}
  \streetaddress{9500 Gilman Dr}
  \city{La Jolla}
  \state{California}
  \country{USA}
  \postcode{92093}
}

\author{Peiling Jiang}
\email{peiling@ucsd.edu}
\affiliation{%
  \institution{University of California San Diego}
  \streetaddress{9500 Gilman Dr}
  \city{La Jolla}
  \state{California}
  \country{USA}
  \postcode{92093}
}

\author{William Duan}
\email{widuan@ucsd.edu}
\affiliation{%
  \institution{University of California San Diego}
  \streetaddress{9500 Gilman Dr}
  \city{La Jolla}
  \state{California}
  \country{USA}
  \postcode{92093}
}

\author{Xiaoshuo Yao}
\email{x5yao@ucsd.edu}
\affiliation{%
  \institution{University of California San Diego}
  \streetaddress{9500 Gilman Dr}
  \city{La Jolla}
  \state{California}
  \country{USA}
  \postcode{92093}
}

\renewcommand{\shortauthors}{Xia et al.}

\begin{abstract}
Despite the advances and ubiquity of digital communication media such as videoconferencing and virtual reality, they remain oblivious to the rich intentions expressed by users. Beyond transmitting audio, videos, and messages, we envision digital communication media as proactive facilitators that can provide unobtrusive assistance to enhance communication and collaboration. Informed by the results of a formative study, we propose three key design concepts to explore the systematic integration of intelligence into communication and collaboration, including the panel substrate, language-based intent recognition, and lightweight interaction techniques. We developed CrossTalk, a videoconferencing system that instantiates these concepts, which was found to enable a more fluid and flexible communication and collaboration experience. 

\end{abstract}



\begin{teaserfigure}
  \centering
  \includegraphics[width=0.88\textwidth]{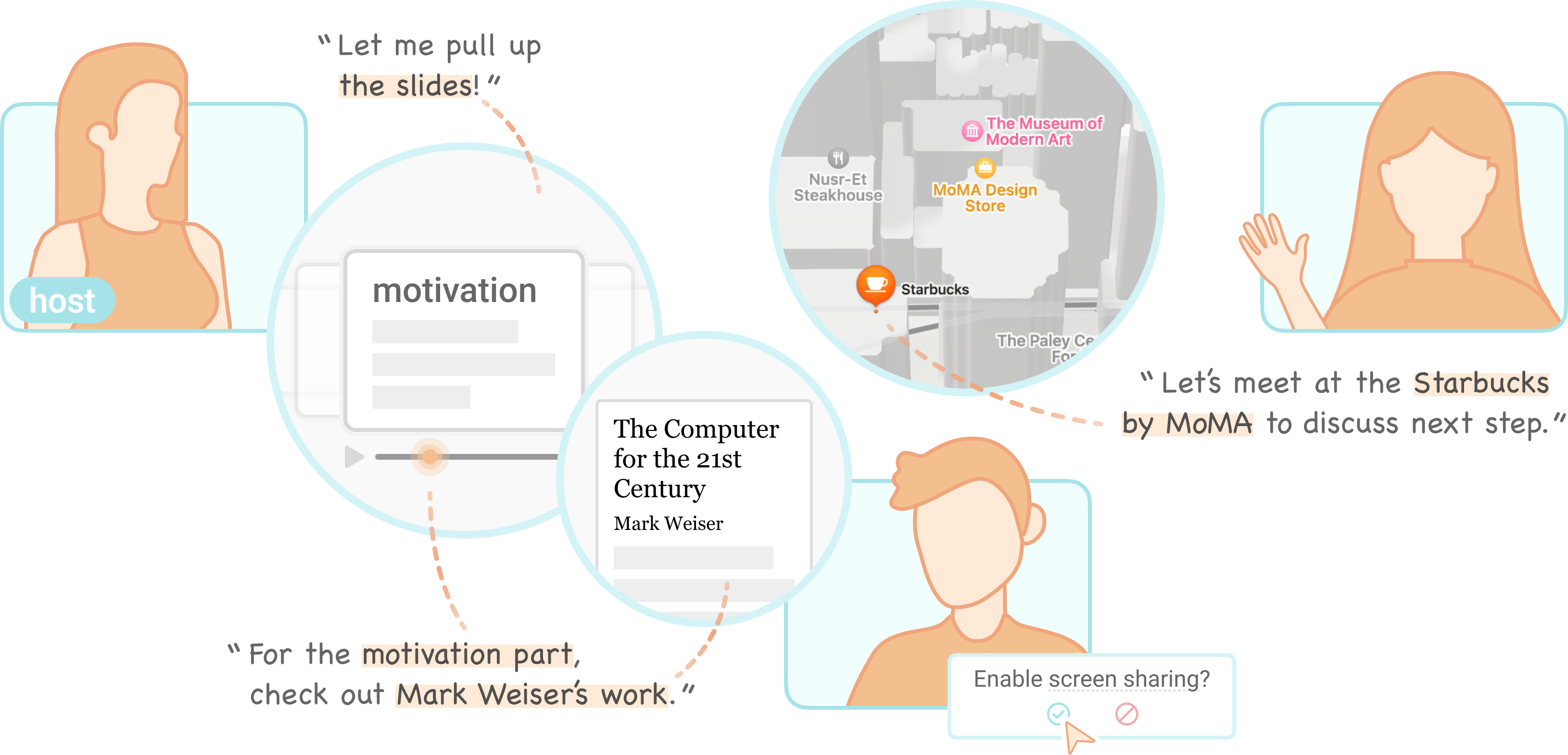}
  \caption{CrossTalk is a video-based communication and collaboration system that infers users’ intentions from conversational speech and provides intelligent, context-aware, and socially acceptable assistance to enhance communication and collaboration.}
  \Description{}
  \label{fig:teaser}
\end{teaserfigure}

\maketitle

\section{Introduction}
Real-time human communication and collaboration such as presentations and discussions, is becoming increasingly moderated by digital communication media. A critical challenge of such media is the lack of shared context amongst the participants, which often leads to ineffective communication and collaboration. To address this challenge, significant research in HCI and CSCW has been conducted to reintroduce contextual cues, such as gaze \cite{ref38Clearboard} and spatial context \cite{ref67BlendedInteraction,ref68Room2room}, to create a sense of \textit{``being there''} \cite{ref35BeyondBeingThere}.

Reintroducing these contextual cues enables videoconferencing participants to gain greater awareness of each other’s context. However, it has long been argued that this approach can never be \textit{``close enough''} in imitating the complexity of real-world interaction \cite{ref35BeyondBeingThere}. Users may also prefer to restrict certain contextual information when using videoconferencing for privacy reasons. On the other hand, prior work has shown that people are often aware of the absence of critical context that may cause miscommunication and proactively provide additional explanations and coordination to inform others of their intent and situation \cite{ref22CoordinationOfCommunication}. For example, when one anticipates a time-consuming information retrieval task, they may inform others by saying \textit{``give me a second, let me dig that file up''}, before or as they are retrieving the file. 

Our insight is that the context and intentions found in these utterances can also be received and interpreted by the communication media to provide intelligent assistance. As such, instead of videoconferencing applications being an oblivious transmission channel of audio and video, we seek to transform them into intelligent, proactive, and socially acceptable participants in video-based communication and collaboration. Inspired by recent research that demonstrated the promise of retrieving and presenting relevant images based on the verbal conversation \cite{ref94Crosscast}, we further explore whether such an approach can be broadly applied to the video-based communication and collaboration environments which include attendees, artifacts, workflows, and the rich interactions among them.

To understand the opportunities for leveraging conversations during videoconferences for intelligent support, we conducted a formative study that analyzed recordings of remote meetings. We found that a considerable amount of verbal coordination is required to establish a common understanding of out-of-context and transient meeting materials, such as links to external content or screensharing sessions that had ended. In addition, significant cognitive demands were placed on users during meetings as they struggled to remain cognizant of ongoing conversation as well as the implicit and explicit requests from other attendees while they are operating the interfaces or engaging with other tasks (e.g., taking notes), further demonstrating the need for the system to offload the cognitive and manual effort for users.  

Informed by these findings, we propose leveraging language-informed intelligence to support communication and collaboration environments with the following components: (1) a persistent substrate to maintain content, context, and actions of meeting elements; (2) an intent recognition and recommendation mechanism that matches user conversations with the content, context, or action of meeting elements; and (3) interaction techniques that enable flexible interaction with the system's recommendations.

We developed a prototype videoconferencing system, CrossTalk, that instantiates these design concepts. As a whole, user study participants found CrossTalk reduced the cognitive and manual effort needed to maintain the flow of conversation and collaboration, enabling fluid and flexible communication and collaboration experiences. As the panel substrate and natural language are both highly versatile, we believe the proposed concepts and techniques can be broadly applied to other communication media, such as interactive displays or augmented reality, for a variety of contexts.

The contribution of this work includes: 
\begin{enumerate}
    \item Findings from a formative study uncovering challenges during communication and collaboration with videoconferencing systems and opportunities to employ conversational speech to provide intelligent and context-aware assistance.
    \item Holistic design of panel substrates, language-based intent recognition, and interaction techniques that enable intelligent and fluid communication and collaboration.
    \item CrossTalk, a prototype video-based communication and collaboration system instantiating our design concepts, evaluated with technical and user evaluations. 
\end{enumerate}

\section{Related Work}
This research draws on prior work on understanding the challenges of online meetings, computer-mediated communication and collaboration, natural language interfaces, and interactive substrates. 

\subsection{Effective Communication and Collaboration}
Research in HCI and Management Science has investigated the challenges that exist during group communication with a significant interest in workplace meetings \cite{ref19MeetingEffectiveness,ref21TheorizingPractice,ref25ATeamCollaborationSpace}. In particular, it was found that the use of meeting artifacts impacts meeting effectiveness \cite{ref59ReliabilityAndInterRaterReliability,ref66ImpactOfMeetingProcedures} and information exchange across participants, groups, and their organizations \cite{ref20Informationrichness}. Although artifacts, such as agendas and summaries, improve meeting effectiveness \cite{ref19MeetingEffectiveness}, they require significant labor to produce before, during, and after a meeting \cite{ref43ImprovingRemoteCollaboration}. Tasks such as note taking induce a high cognitive load for meeting participants as they need to engage with multiple tasks simultaneously \cite{ref39Dynamo,ref41HumanConversationtrainingdata}. Due to the high cognitive load, a dedicated facilitator is often required to ensure that a meeting follows an agenda and important information is recorded \cite{ref73WastedTimeAndMoneyInMeetings,ref87TheGroupFacilitator}.

Recently, videoconferencing has become an essential part of modern society and has become increasingly integrated into our personal lives \cite{ref3Thehyperpersonal,ref6Canvideoconferencingaffect}, work \cite{ref32PenTouch,ref44MeetingsMatter}, and education \cite{ref10Radiologyeducation,ref54VisualCaptions,ref89ContextAwareIntentIdentification}. While videoconferencing technologies enable face-to-face communication over a distance \cite{ref34Distributedwork}, the challenges that exist during collocated meetings transfer to this new medium and become further exacerbated by the lack of eye contact, body language, spatial references, artifacts, and implicit social cues \cite{ref27AGlobalOptimization,ref53VideoInInteraction,ref58WhyCscwApplicationsFailProblemsInTheAdoptionOfInterdependentWorkTools}.

While the present research falls under the umbrella of supporting effective meetings, it focuses on the friction that users encounter while communicating and collaborating during videoconferences. By analyzing a set of remote meetings, a list of friction points was identified, many of which also exist during collocated meetings \cite{ref43ImprovingRemoteCollaboration}. In this work, we employ intent recognition to infer and automate users’ intents to reduce the friction as well as identify opportunities to further augment their meeting experiences.

\subsection{Computer-Mediated Communication}

Significant work has explored how to improve computer-mediated communication so that it can afford the same richness and variety of contextual, non-verbal cues and interactions that occur when attendees are collocated. ClearBoard preserved the gaze awareness of remote users by integrating remote video feeds with a shared workspace \cite{ref38Clearboard}. Room2Room enabled for the life-size telepresence of remote users via projected augmented reality \cite{ref68Room2room}. MirrorBlender enabled users to position video views to achieve spatial consistency amongst each other to support the use of nonverbal cues (e.g., deictic gestures) in videoconferences \cite{ref29Mirrorblender}.

Recent research has started to leverage computer-mediated communication media as interaction mechanisms that are valuable beyond replicating the physical world to go \textit{``beyond being there''} \cite{ref35BeyondBeingThere}. For example, SnapStream demonstrated that snapshots of live video could be used to enhance viewer engagement and commentaries during art live streams \cite{ref95Snapstream} and LiveMâché illustrated how contextual sharing improved participation for livestream viewers in online learning settings \cite{ref31CollaborativeLiveMediaCuration}. 

Other systems have aimed to build collaborative spaces for group interaction. MeetCues implemented and visualized casual interactions to improve engagement in online meetings \cite{ref4MeetcuesSupportingOnlineMeetingsExperience}. WeSearch, for example, enabled a group of users to conduct collaborative web searches on a large tabletop display \cite{ref65Wesearch}. Dynamo enabled users to gather, share, and exchange information during meetings that took place around public communal surfaces \cite{ref39Dynamo}. Although these systems have shown considerable promise in facilitating communication, they still require users to manipulate interfaces while talking \cite{ref83MeerkatAndPeriscope}, which places cognitive strain on users. Crosstalk, on the other hand, aims to automate users' intended tasks, reducing the mental and manual effort required from them.

\subsection{Natural Language Interfaces}
Natural language interfaces have seen use in a variety of domains, such as data visualization \cite{ref22CoordinationOfCommunication,ref82OrkoFacilitatingMultimodal} and image editing \cite{ref51Pixeltone}. Pioneering systems such as Put That There \cite{ref11PutThatThereVoiceAndGestureAtTheGraphicsInterface} and SHRDLU \cite{ref90UnderstandingNaturalLanguage}, and voice-based assistants use explicit natural language commands as input to interact with computers \cite{ref1Siri,ref57AugmentingConversations,ref61AzureCognitiveServices}. 
Later, interfaces and systems that infer desired information and actions from natural language expressions not intended as command input became a topic of interest \cite{ref14Messageontap,ref89ContextAwareIntentIdentification, ref79Meetingvis, ref91Crosspower}. For example, CrossPower leveraged video scripts to infer desired visual imagery and graphic effects \cite{ref91Crosspower}. CrossCast \cite{ref94Crosscast} and Visual Captions \cite{ref54VisualCaptions} enabled the retrieval of relevant images as visual aids based on user conversations to augment verbal communication. MessageOnTap explored a suggestive interface based on the text in a messaging application to recommend relevant content and interface actions \cite{ref14Messageontap}.

CrossTalk infers users' intentions from their conversations and recommends relevant content and actions, and thus, is a type of natural language interface. Different from many prior works, CrossTalk seeks to augment real-time synchronous human communication, requiring recommendations and interactions to cause minimum disruptions to the flow of communication. To achieve this, CrossTalk follows established guidelines for designing mix-initiative and AI-infused systems by enabling the users to utilize lightweight interaction to control the system's recommendations \cite{ref26PrinciplesOfMixed, aidesignguidelines}.

\subsection{Information Objects and Substrates}
CrossTalk builds upon prior work that explores the persistent, object-oriented representation of information for dynamic and intelligent actions \cite{beaudouin2017towards, webstrates, ref92ObjectOrientedDrawing, ref93Spacetime, xia2020object}. For example, Xia et al. proposed the object-oriented interaction paradigm, which explores the reification of information and structures as interactive objects that be flexibly manipulated and composed to support higher-order tasks \cite{xia2020object, ref92ObjectOrientedDrawing, ref93Spacetime,dataink}. Recent work by Klokmose and Beaudouin-Lafon on information substrates also demonstrates that reifying abstract data as interactive and interoperable objects enables composability and cross-platform collaboration \cite{webstrates, vistrates, codestrates}. For example, Webstrates builds upon the concept of shareable dynamic media \cite{PersonalDynamicMedia} as a framework for representing content in a shareable and persistent manner that allowed for cross-device collaborative interaction \cite{webstrates}.

The object-oriented paradigm and the concept of information substrates embody information as objects, laying the foundation for enabling objects to preserve context and exhibit intelligent behaviors. For example, in Object-Oriented Drawing, each Attribute Object retains its history for attribute-level undo/redo, circumventing the cost of losing operations with application-level undo/redo \cite{ref92ObjectOrientedDrawing}. Closely related, in Spacetime, a VR-based collaborative scene editing environment, selections of objects are reified as Container Objects, which preserves the selection location to enable quick navigation to the original design context \cite{ref93Spacetime}.

CrossTalk further explores language-informed intelligent interaction with information. By encapsulating content, context, and actions within each object shared in the videoconference, and matching them with real-time user conversations, CrossTalk is able to recommend relevant information and interface actions to the users. This shifts the manual and cognitive effort from the users to the intelligent communication media, delivering an enhanced video communication and collaboration experience.

\section{Formative Study}

To identify sources of friction and opportunities to support video-based communication and collaboration, a formative study was conducted to analyze recordings of Zoom meetings. We collected a dataset of recorded meetings representing a variety of previously reported typologies of meetings \cite{ref4MeetcuesSupportingOnlineMeetingsExperience,ref25ATeamCollaborationSpace,ref60MoreToMeetings}. For analysis, we employed reflexive thematic analysis \cite{ref12Reflectingonreflexivethematicanalysis} to identify patterns in the data with three coders meeting to reflect on the codes and seek agreement during each round of coding. Codes and themes were generated inductively starting with an initial shortlist of lean codes that were informed by the coders’ prior review of literature on meeting software and language interaction (e.g. \cite{ref60MoreToMeetings, ref57AugmentingConversations}). Following iterative theoretical sampling for video analysis \cite{ref48VideographyAnalysingvideodataasafocusedethnographicandhermeneuticalexercise, ref49Videoanalysisandvideography}, new recordings from different sources were added in each round until no new codes emerged due to code saturation and generalizability.

Videos included recordings of meetings where a large amount and variety of information is collected, shared, and discussed. Several recordings were taken from YouTube following fair use guidelines using broad search terms for meetings (e.g. government, business, product). We also added recorded research meetings from our team that were conducted using Zoom during the COVID-19 pandemic as auto-ethnographic research \cite{ref47TracesOfTime}. In total, we gathered 60 hours of recorded meetings of various formats and domains in Appendix~\ref{app:formative}. We did not seek inter-rater reliability or build a hierarchical set of codes following recommended practice \cite{ref59ReliabilityAndInterRaterReliability} as the aim of the formative study is not to produce a theory of behaviors for all meetings, but to inform the design of the CrossTalk system.

\subsection{Informing Actions and Fiction}
During videoconferences, participants were aware of the lack of visibility that other attendees had into their context, which caused confusion and unintended silence. They felt obligated to inform the group about their context, especially when they anticipated that their actions may have taken significant time. For example, participants often verbally expressed their intention (e.g., \textit{``let me find that paper''}) or described the user interface elements they were manipulating (e.g., \textit{``share screen... Chrome... share sound''}). When encountering friction with a user interface, presenters expressed their confusion and frustration (e.g., \textit{``I wish it would just pull up the right file for me''}). Such expressions were often participants’ own think-aloud dialog, but also informed the group of the situation and often resulted in help being provided to them.

\subsection{Lack of Shared and Cognizant Interaction with Materials}
Meeting attendees often verbally referred to materials to direct the group’s attention to upcoming discussion. Because only the presenter who shared their screen had control of the material, attendees’ interaction with the materials had to be delegated to the presenter. If presenters were not cognizant of the desired interaction or occupied by other tasks (e.g., taking notes), other attendees had to explicitly repeat their requests, resulting in communication costs and social pressure. For example, when a participant said, \textit{``I have feedback for the motivation [section of the presentation]''}, the presenter was expected to navigate the screen to the referred slide to provide context. If they failed to do so, the participant would explicitly say, \textit{``can you go back to the motivation slide?''}. The presenter apologized for missing the request and then spent time manipulating the interface.

\subsection{Lack of Meaningful Representations of Meeting Materials}
Due to the nature of the meetings, a variety of materials were shared among participants, including presentations and demonstrations, research articles, images, sketches, messages, websites, and more. We found that the friction mentioned above was often due to the lack of meaningful representations and structures of the materials when shared via screen sharing or as URLs. When sharing their screen, participants opted to share application windows rather than their entire screen for privacy reasons. However, because information related to one research activity often resided in different applications, participants had to repeatedly share and un-share different applications while locating the referred content in different applications, which led to frequent interruptions of the conversations. The use of external links also encouraged individual exploration over collaborative examination. It was often unclear if participants had viewed the material or which part they had viewed. As a result, additional coordination was often required to establish a common context for discussion.

\subsection{Summary}
Our findings show that meeting materials lack appropriate and shared representations. They are transient, localized, and time-consuming to retrieve, leading to communication costs. In addition, due to the lack of visibility into each other’s context, attendees verbally expressed contextual cues to establish a common frame for discussion, requested manual control of materials, and thought out loud about their actions or confusion with the interface in order to avoid silence and increase group awareness of their status. This is consistent with Fussell et al.’s work showing that lack of context and shared materials lead to more explicit verbal descriptions of one’s internal state and task status during communication \cite{ref22CoordinationOfCommunication,ref50VisualInformation}.
 
\section{DESIGN CONCEPTS}

To approach the systematic integration of language-informed intelligence, meeting materials and interface elements need to be upgraded to be able to infer user intention from the conversation as well as present the inference to the user for confirmation and interaction. We propose the following concepts. 

\textbf{C1: Material-Specific Intelligence Description}
 A common approach to user intent inference is to equip a system with a
centralized intent recognition module to recognize predefined intentions. This approach is insufficient, however, to support a dynamic environment where customized materials
of diverse formats can be flexibly included.
To achieve de-centralized intelligence and ensure low barrier for
developers to specify the intelligence, we propose using natural language to describe the content, context, and tools within each panel, which can be matched with users’ real-time conversations
to determine relevance.

\textbf{C2: Intent Recognition by Matching Users' Conversations with Material-Specific Descriptions}
User intents can be recognized by associating users’ conversations with the descriptions of content, context, and actions of materials within an environment. This can be achieved by computing the semantic similarity of a natural language description with the conversation context. Given the advent of large-language models, one can imagine the description of a tool can be defined as a prompt to an LLM-based intent recognition engine for advanced operations such as parameter extraction. For example, given a map, the navigation function can be augmented with a prompt to extract all mentioned locations (e.g., \textit{``extract all location entities into an array from the input text''})

\textbf{C3: Visualization of System Inferences and Ignorable, Reversible, and Lightweight Interaction}
Uncertain and erroneous user intention inferences are inevitable. Therefore, the user interface should clearly communicate the system’s status and enable users to easily accept, reject, or ignore the system inference, as well as easily recover from errors to avoid introducing additional friction during communication and collaboration. \\

\begin{figure}
  \includegraphics[width=\linewidth]{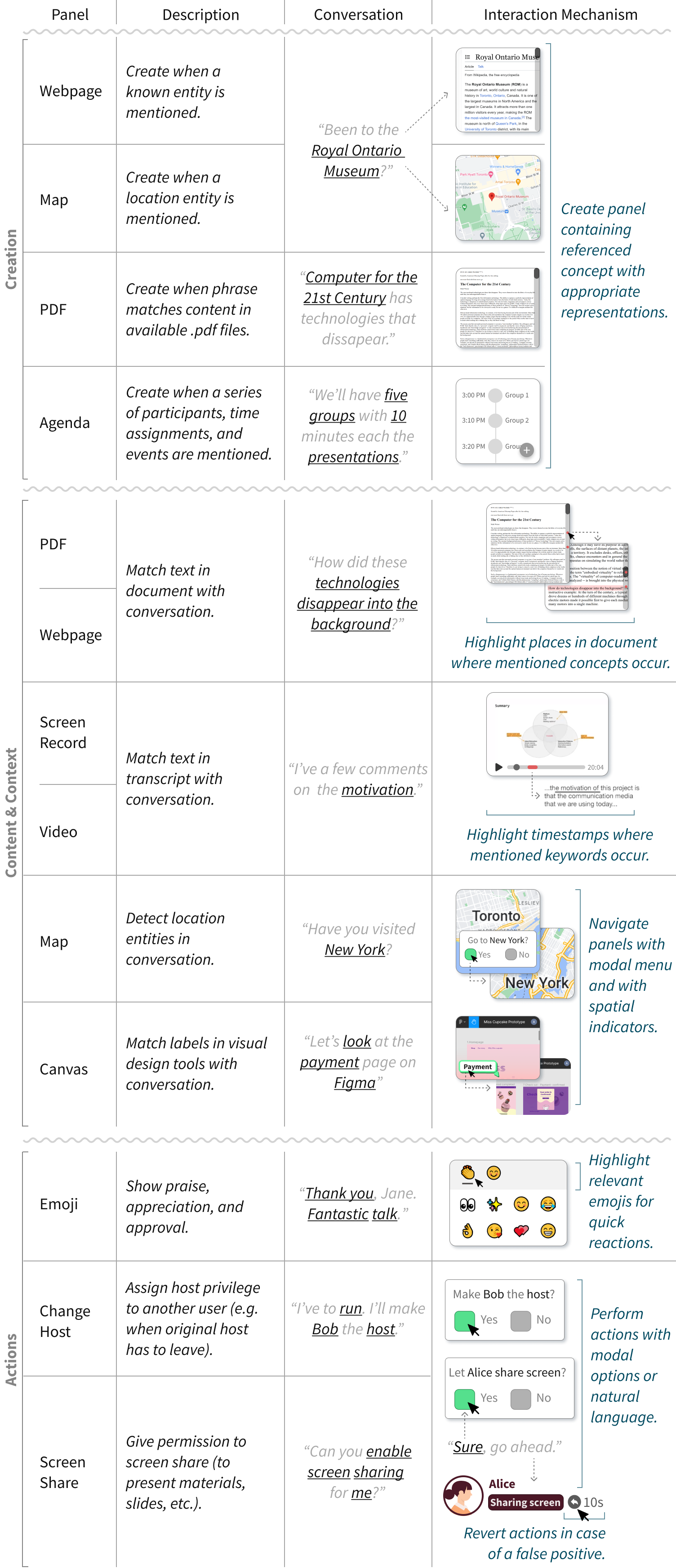}
  \caption{The Design Space and Examples of the Panel-specific Natural Language Description, Conversational Speech, and the Interaction Mechanisms.}
  \Description{}
  \label{fig:interaction-mechanism}
\end{figure}

Combining these concepts, we propose (1) a panel substrate that maintains the content, context, and actions of digital elements as well as their descriptions; (2) an intent recognition engine that matches users' conversational speech with all present panels and their content, context, and action; (3) a consistent interaction design vocabulary to facilitate the interaction with system recommendations. Figure \ref{fig:interaction-mechanism} shows a design space based on these concepts.

\section{CrossTalk}
We designed a proof-of-concept videoconferencing and collaboration system, CrossTalk, that instantiates the above concept to investigate their feasibility.

\begin{figure*}[ht]
  \includegraphics[width=\textwidth]{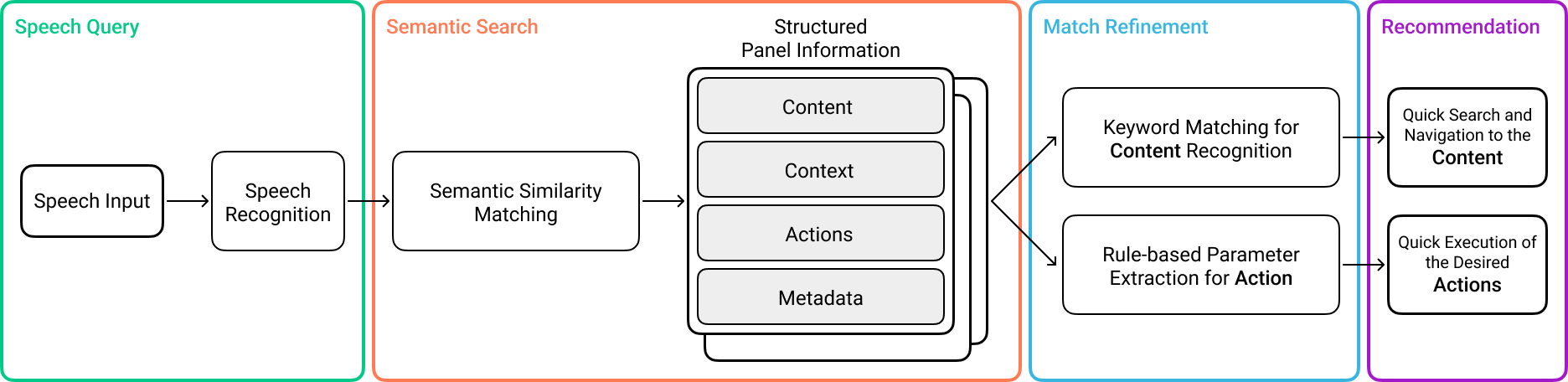}
  \caption{User Intent Recognition Pipeline.}
  \Description{}
  \label{fig:intent-pipeline}
\end{figure*}

\subsection{Panel Substrates}
To provide a suitable substrate for intelligent assistance, we propose extending panel designs within existing videoconferencing user interfaces to not only show a list of meeting attendees, but also all relevant meeting artifacts and materials. This should enable all meeting resources, such as video feeds of attendees, screenshared content, materials, and meeting artifacts, to be represented with a consistent style. Each panel houses an independent unit of information along with corresponding functionality. Together, they aggregate relevant meeting resources, allowing participants to interact with meeting materials in situ.

\subsubsection{Panel Types} CrossTalk supports three types of panels. Attendee Panels house meeting attendees’ audio and video feeds and are created when attendees join the meeting. A transcript panel shows the real-time transcript of attendees' conversations. Content Panels contain shared meeting materials, including maps, images, videos, whiteboards, PDF documents, search engine results, webpages, polls, and agendas.

\subsubsection{Panel Manipulation} Panels can be flexibly created, manipulated, stored, or removed. Attendee Panels can be closed by the attendee themselves to leave a meeting. Content Panels can be minimized to the side when no longer needed. Users can also directly share panels to the storage area to avoid distracting the group and or cluttering the canvas. Content Panels can be prepared and shared before the meeting so that they are accessible to attendees when the meeting starts. Panels can enter a presentation mode to gain the attention of meeting attendees, which is equivalent to screensharing in videoconferencing systems. The difference, however, is that a screensharing session is automatically recorded as a private video panel, which can be made public for discussion.

\subsubsection{Panel Actions, Content, and Context for Intent Recognition} To enable panel-based intent recognition, each panel maintains all the relevant metadata, actions, content, and context for intent recognition. For example, the metadata includes the type, creator, time, and access information of a panel. The actions of a panel contain the supported actions as well as their descriptions which can be used to compute semantic relevance. The content of a panel consists of all the recognizable content, such as text, images, and videos. The context records all the past interactions with a panel, including views, searches, and conversations about the panel.

\subsection{User Intent Recognition}
Given users’ conversational speech as input, our goal is to infer user intent and recommend relevant panels, content, and actions to users so that they can quickly navigate to desired content or execute desired actions. CrossTalk employs a 4-step pipeline that consists of semantic search and rule-based mechanisms to extract desired content and actions from conversational speech (Figure \ref{fig:intent-pipeline}). It (1) converts user speech to text and uses the text as queries; (2) searches all the panels, their content, context, and actions, based on their semantic relevance to the search query; (3) refines the most relevant content, context, and actions identified from the previous steps to extract the relevant parameters of the actions and provide precise content matches; (4) recommends top-scoring panel content and actions so that users can quickly navigate to desired content or execute desired actions.

\subsubsection{Matching Content and Actions from Conversation Speech}

As a user speaks, CrossTalk continuously recognizes their speech. CrossTalk employs a sliding window technique and uses the 10 mostly recently recognized words as the search context.  CrossTalk leverages a BERT-based semantic search to find the actions and content in the panels that are most relevant to the query \cite{ref71SentenceBert}. The principle of semantic search is to find units of semantically similar language in a corpus by calculating the semantic distance between the query and the units (commonly sentences) in the corpus in the vector space \cite{ref62Cortana}. The key benefit of this approach is that it can recognize similar semantics from the diverse expressions and vocabularies users employ to refer to the same interface action and content. With CrossTalk, all the text within panels (e.g., text from PDFs, image captions, video transcripts, descriptions of interface actions, etc.) was used as the corpus, and the search query was encoded as a vector to find the most semantically similar actions or content from the panels. 

\begin{figure*}
  \includegraphics[width=\textwidth]{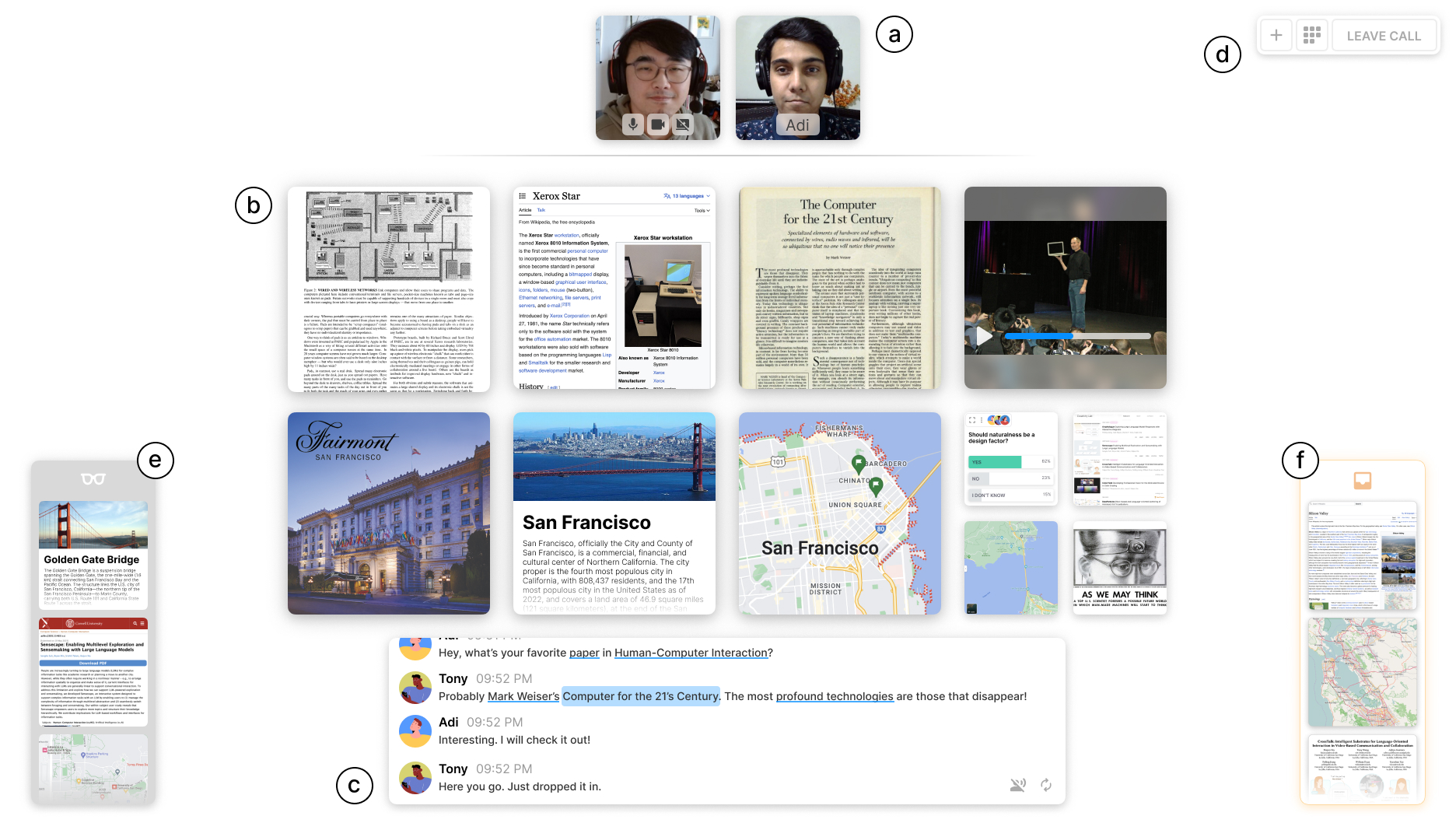}
  \caption{CrossTalk Interface. a) an attendee panel, b) the content panel, c) the transcript panel which shows recognized text from speech, d) user interface control for joining the call, e) private panel area that is only visible to meeting attendee, f) the storage area for minimized panels.}
  \Description{}
  \label{fig:crosstalk-interface}
\end{figure*}

\subsubsection{Refining Content and Action Searches} Matching conversations with content and actions on the sentence level enables CrossTalk to take advantage of the context of sentences for an accurate semantic search. While it is adequate for coarse content search, it is insufficient for recommendations of interface actions, as they may require additional parameters as well as more precise information searching (e.g., pinpointing a timestamp in a video).

To address these issues, CrossTalk further refines content and action search using a rule-based mechanism, commonly used in task-oriented conversational agents \cite{ref40ChatbotsAndDialogue}. Our customized rules are based on a set of pre-defined verbs, nouns, and preposition phrases derived from the formative study, which are mapped to corresponding interface functionality. Depending on the type of content and action, relevant elements were extracted from the text to form a complete interface action or provide fine-grained content matches. For example, for a map-based navigation action, CrossTalk would identify location entities from a search query as parameters of the navigation; if a sentence in a video transcript was identified as the most relevant sentence, CrossTalk would further match entities from the search query with those from the search results based on their semantic similarity to return fine-grained timestamps.

\subsection{Implementation}
CrossTalk was developed as a web application using React as the front-end framework, Agora for real-time video communication, and Firebase for storing and synchronizing content across multiple users. The intent recognition pipeline consists of speech recognition using Web Speech API \cite{ref16GroundingInCommunication}, semantic search using the BERT-based Transformers \cite{ref71SentenceBert,ref76SentenceTransformers}, and Google Cloud NLP \cite{ref28GoogleNLP} for entity recognition required by the fine-grain intention recognition. 

\subsection{Summary}
The panel substrates and intent recognition provide high composability to allow CrossTalk to host information in diverse formats with panel-specific intelligent recognition. When a panel is added, it provides all of its content and actions to CrossTalk, enabling CrossTalk to perform intent recognition within the context of the entire information environment. This enables future panel developers to define their own panel-specific intelligent rules that are suitable for their specific needs and goals.

\section{INTERFACE AND INTERACTION}

When designing the interface and interactions that leverage inferred user intents, careful consideration of the potential issues the system may face, including speech recognition errors and false detection of user intents, was taken into account by following established guidelines for designing mixed-initiative \cite{ref26PrinciplesOfMixed} and AI-infused systems \cite{aidesignguidelines}.

The CrossTalk interface consists of four main areas (Figure \ref{fig:crosstalk-interface}). In the center is the canvas on which the panels are displayed. A transcript panel at the bottom that shows real-time speech recognition can be minimized to yield more space for the main canvas. A private area hosts content panels that are private for the user. A storage area stores minimized content panels for the ongoing session or collections of panels from previous sessions.

\begin{figure}
  \includegraphics[width=\linewidth]{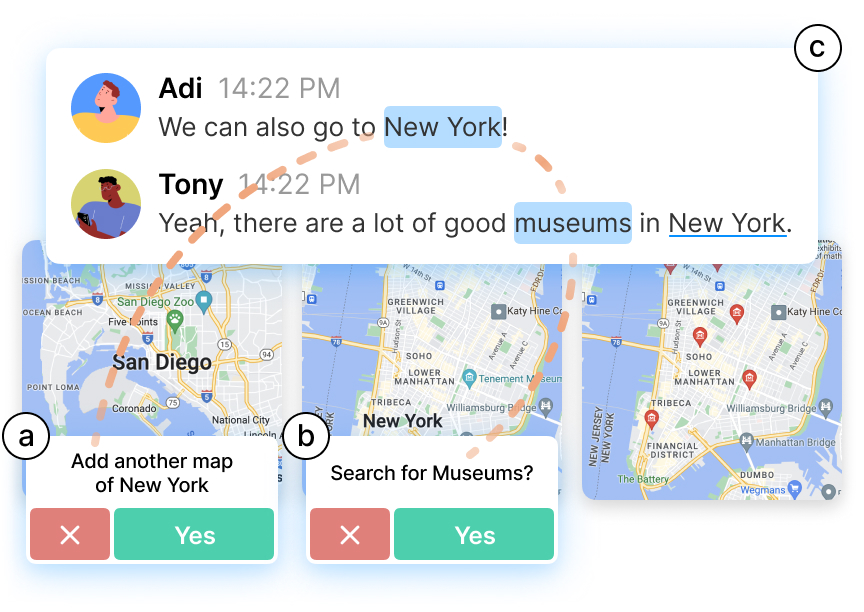}
  \caption{CrossTalk generates prompts to enable users to quickly navigate the map (a, b). The user can accept the prompt to adjust the search in the corresponding view (c).}
  \Description{}
  \label{fig:generates-prompts}
\end{figure}

\subsection{Panel Creation and Basic Manipulation}
Content Panels are recommended when potential entities of interest are detected in speech and used as search queries for relevant information (e.g., map, webpages, video). In CrossTalk, Content Panels are not directly recommended on the canvas to avoid distracting users. As there can be many entities mentioned in a short period of time, the search queries may need further adjustments to retrieve relevant information, or there are multiple possible content formats of the same query (e.g., map vs. Wikipedia vs. general search results for a location). Therefore, the recommendations are visualized in the Transcript Panel. Users can flexibly adjust their queries via text selection and choose the desired visual forms before adding the panel to the canvas (Figure \ref{fig:generates-prompts}). When the Transcript Panel is minimized, the entities recognized from the speech are displayed as a slowly flowing stream of minimized panels. The user can drag and drop a minimized panel to create a content panel and place it at a desired location on the canvas. Alternatively, Content Panels can be created manually by dragging and dropping materials into the system or by copying external links into CrossTalk. 

Persistent representation of the panel enables rich awareness and interactions with previously transient information. Like most videoconferencing applications, CrossTalk enables screen sharing, which is automatically recorded when initiated. Once the screensharing session is over, the presenter can make the recording public to facilitate the continued review and discussion of the recording.

\begin{figure}
  \includegraphics[width=\linewidth]{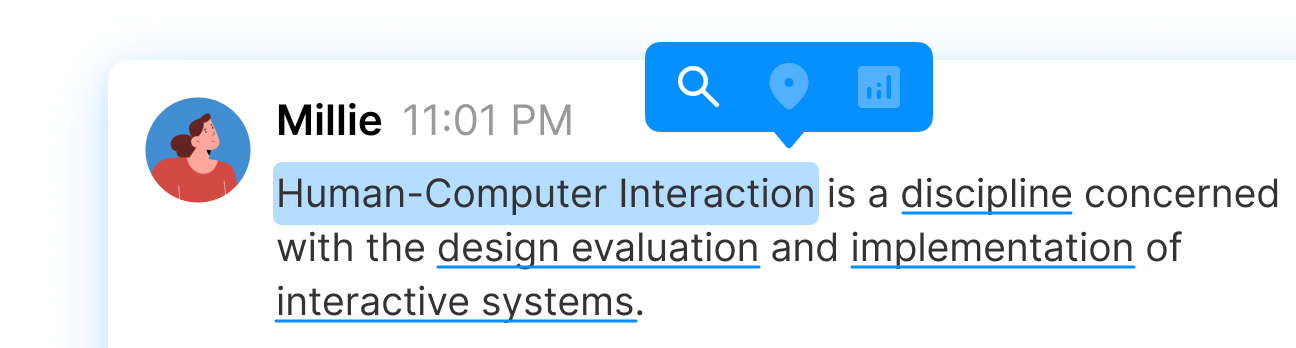}
  \caption{Recognized entities are highlighted, and the user can select any text to refine the query.}
  \Description{}
  \label{fig:recognized-entities}
\end{figure}

\subsection{Panel Organization}
Panels can be organized in a variety of ways to suit diverse needs, such as in a grid or in canvas layouts. A challenge when employing a panel substrate is that it may lead to visual clutter when a large amount of information is shared. To address this problem, CrossTalk dynamically adjusts the size of panels based on the last time they were interacted with and the usage frequency of the panels. For example, when a panel has not been referenced or interacted with, it gradually shrinks in size to give space to other panels. When they reach a certain size threshold, they will be moved from the main canvas to the storage area.

\subsection{Interaction with the System Inference}
CrossTalk enables users to interact with the system inferences using lightweight user interface widgets and interaction techniques. 

\subsubsection{Panel Content Recommendation} Attendees often need to navigate to a section of a document or video as context during a discussion. However, manual navigation within a large file is time-consuming and interrupts the flow of communication and collaboration. Whenever CrossTalk detects that there is content relevant to the user conversation (e.g., text in PDFs, webpages, video transcripts), it highlights the corresponding sections and provides navigation shortcuts on the sliders of documents and the timelines of videos. When multiple matches are available, CrossTalk encodes the semantic similarity of the recommendations via the color intensity of the navigation shortcuts. When users hover their cursors on these navigation shortcuts, CrossTalk reveals the key matching entities and enables users to preview the search results before navigating to the recommended section or clicking on the navigation shortcuts to take them directly to the sections.

\begin{figure}
  \includegraphics[width=\linewidth]{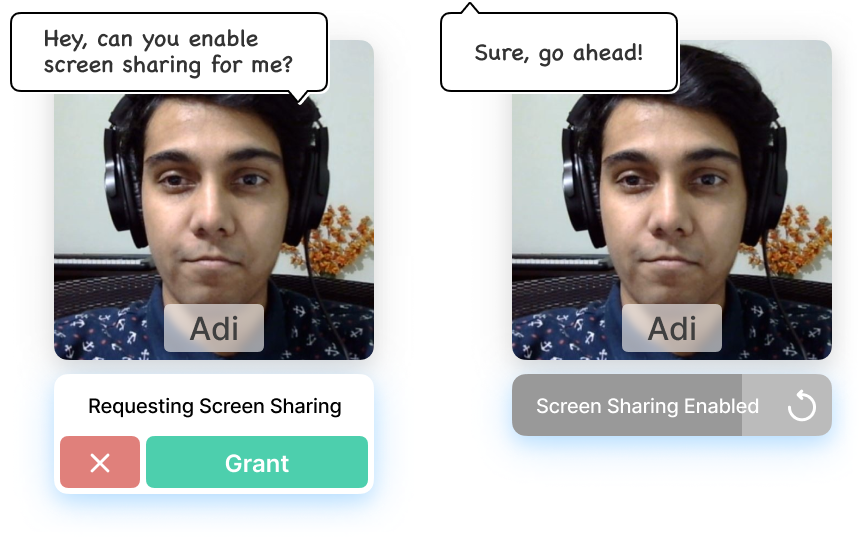}
  \caption{Left: CrossTalk detects an attendee is requesting screen sharing and prompts the interface action. Right: permission is granted verbally, and a revert button is prompted in case of any error.}
  \Description{}
  \label{fig:request-screenshare}
\end{figure}

\subsubsection{Panel Action Recommendation} When interface actions are inferred, CrossTalk shows the actions in the corresponding panels and prompts users to accept or reject the recommendations. For example, if a participant says, \textit{``we can also go to New York''}, a prompt will appear on top of an existing map panel on the canvas, allowing the user to quickly navigate to the location (Figure \ref{fig:generates-prompts}a). However, the user may wish to create a new panel alongside the existing one to preserve the original context. They can drag the prompted search query and drop it on the canvas to form a new panel. When the following sentence is detected, \textit{``there are a lot of good museums in New York''}, a similar prompt will appear on top of the New York panel, allowing the user to search museums within the map panel (Figure \ref{fig:generates-prompts}b).

\subsubsection{Verbal Recommendation Confirmation} In some cases, it is natural for a user to respond to others’ requests verbally. CrossTalk leverages these verbal utterances to respond to the system’s action recommendation. For example, when a user speaks \textit{``Can I have screen sharing permission?''}, CrossTalk would notify the meeting host asking whether screen sharing permission has been granted and the host can acknowledge the request verbally (e.g., \textit{``yes, go ahead''}, \textit{``Sure''}) to accept the action recommendation. Because verbal expressions can be misinterpreted, CrossTalk presents a follow-up revert button to allow the user to quickly undo the action (Figure \ref{fig:request-screenshare}). All prompts can be dismissed by the user manually or will disappear after a time threshold (e.g., 5 seconds).

\subsection{Summary}
The generic panel substrate and natural language interactions allow CrossTalk to be applied to a variety of communication and collaboration settings that have various types of information and associated actions. The current CrossTalk prototype was designed to explore interface and interaction mechanisms rather than to address the specific needs of a particular communication and collaboration setting. We anticipate adaption and adjustment will be needed when applying the interaction concepts to different settings, such as different panel layout mechanisms and domain-specific vocabularies. We describe several application scenarios CrossTalk currently supports below.

\section{Application Scenarios}

In this section, we describe experiences and workflows that CrossTalk enables for online teaching and learning, formal meeting management, and presentations and discussions during research meetings.

\subsection{Online Teaching and Learning}
\begin{figure}
  \includegraphics[width=\linewidth]{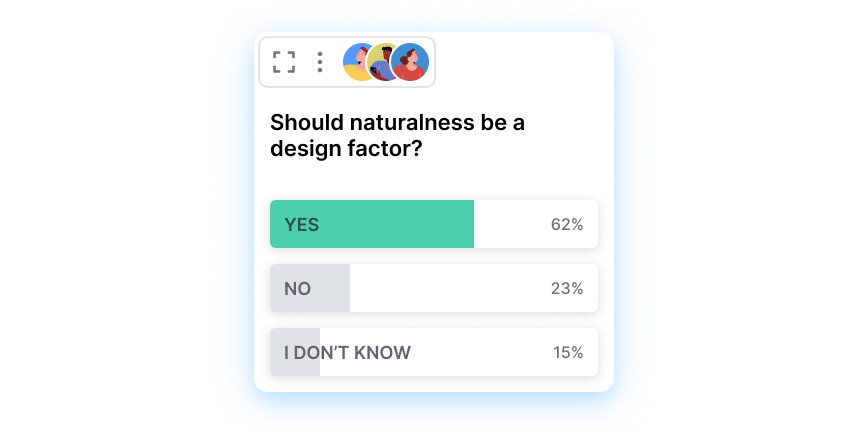}
  \caption{A poll panel generated by \textit{``My question is should naturalness be a design factor?''}.}
  \Description{}
  \label{fig:poll}
\end{figure}

CrossTalk provides several panels with language-oriented interaction techniques to facilitate online teaching and learning. For example, polls are often used to engage students with lecture materials. However, due to the considerable manual effort needed to create polls, instructors must prepare them beforehand, which prevents spontaneous engagement during class. With CrossTalk, the verbal expression of a question, such as \textit{``Here is a question: should naturalness be a design factor?''} can be recognized as a poll panel shared with all attendees to collect responses (Figure \ref{fig:poll}).

\begin{figure}[t]
  \includegraphics[width=\linewidth]{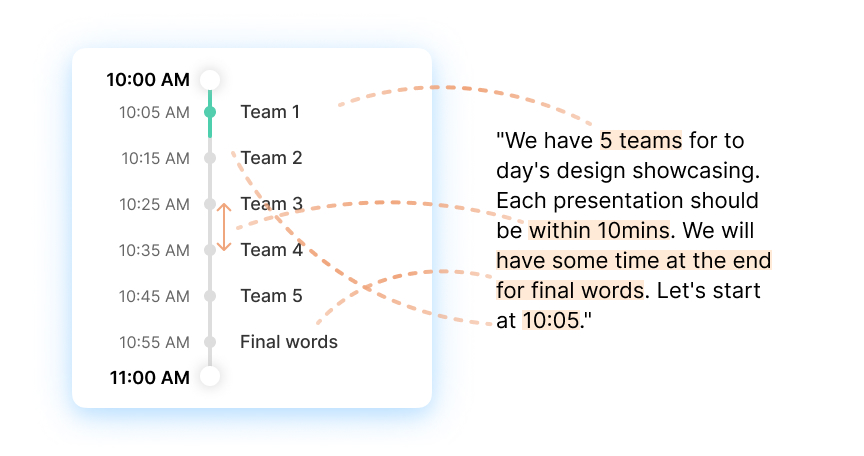}
  \caption{An agenda panel generated by the text on the right.}
  \Description{}
  \label{fig:agenda}
\end{figure}

Keeping track of the time during classroom presentations is often cognitively demanding. Instructors need to constantly check the time, make sure it adheres to the agenda, and use appropriate social protocols to adjust the flow, such as interrupting the current presentation. CrossTalk can detect instructors' verbal descriptions of the order and time duration of the presentations, and generate a populated agenda panel (Figure \ref{fig:agenda}). The instructor can modify the proposed agenda and share it publicly with the students. As the time approaches the end of an agenda item, a prompt will be displayed on the agenda panel to remind students of the time.

\begin{figure*}
  \includegraphics[width=\textwidth]{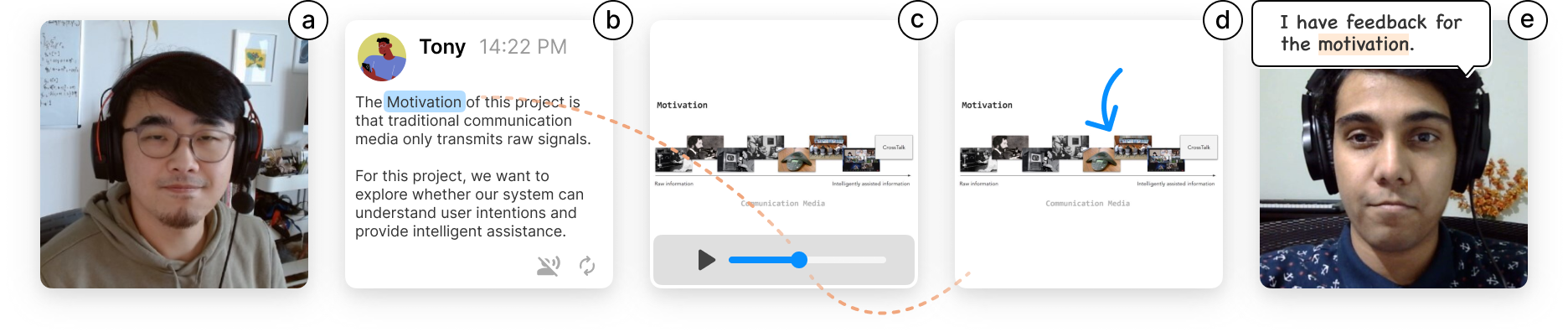}
  \caption{Discussion over screenshared content. After speaker (a) finishes the presentation, he can share the recorded screensharing video with the group.  Meeting attendee (e) engages in the discussion by saying \textit{``I have feedback for the motivation.''} CrossTalk interprets the referred context and highlights the corresponding timestamp of the video, (c) based on the captured speech during the presentation (b). The attendee (e) drags the desired frame out and annotates it while providing feedback.}
  \Description{}
  \label{fig:screenshared-content}
\end{figure*}

\subsection{Formal Management Meetings}
The poll and agenda panels can also facilitate the flow of formal government and management meetings for a diverse range of organizations, such as governments, associations, unions, and boards. Many formal meetings have adopted the format and guidelines introduced by Robert’s Rules of Order \cite{ref72Robertsrulesoforder}. In such meetings, procedures are managed with a specialized vocabulary, with phrases including call to order, second, vote, and yield, each of which signifies a corresponding action for the meeting. 

The Robert’s Rules specify the formal definitions, pre-defined expressions, and corresponding actions of the rules. For example, to amend a motion, a meeting attendee needs to say \textit{``I move that this mention be amended by ...''}, which needs a second and majority vote. In CrossTalk, the Robert Rules can be specified as interface actions for it to recognize the desired meeting actions and provide intelligent support for meeting attendees such as generating a vote for a motion, notifying meeting attendees for a required second, or intelligently muting speakers when the speaker yields the floor.

Formal meetings also utilize pre-determined agendas to ensure effective discussion. With CrossTalk, agenda panels can be created from written or verbal descriptions. Alternatively, a meeting host can log in to the meeting before it starts to define the agenda in place, rather than preparing it in a separate document.

\subsection{Presentations and Discussions}
CrossTalk can also facilitate discussions commonly seen in team collaboration. When a user presents via screen sharing, the screen-shared content is automatically recorded as a persistent video.  When the presenter stops screen sharing, the video is saved in the private area on the presenter’s side. The presenter can drag the video from the private area to the canvas to make it accessible to meeting attendees, which allows them to flexibly navigate, review, and comment on the content. Because every panel also maintains relevant context, in this case, the video panel records the speech content while it was presented. This enables intelligent speech-driven video navigation, as CrossTalk matches ongoing conversations (e.g., \textit{``I have feedback for the motivation''}) with the retained context and recommends relevant timestamps in the video. The user can easily navigate to a desired position by clicking on the recommended timestamps. Thus, the effort to search and navigate to a specific context is mitigated for both the presenter and the feedback-giver. Instead of directly annotating on top of a public video panel, which may prevent others from reviewing the content it depicts, a user can drag the frame of the video away as a separate image panel for detailed visual annotation (Figure \ref{fig:screenshared-content}).

\section{Technical Evaluation}
We conducted a technical evaluation to assess the performance of the user intent recognition in our current prototype. We report on the experiment settings, results, and potential improvements.

\subsection{Experiment Settings}

\subsubsection{Recommendation Confirmation}
The CrossTalk system builds upon BERT-based semantic search technique, which can perform sentence similarity computation at 14200 times per second \cite{ref71SentenceBert}. Since the content search is a direct employment of this technique, which has been extensively evaluated in the NLP literature, we focused on evaluating the performance of the action recognition. Specifically, we gathered a dataset of sentences, each of which was labeled with the interface action it may indicate. We then input each sentence in the corpus into the user intention recognition module and compared the suggested actions with the ground truth.

\subsubsection{Dataset and Action Recognition Model}
While we sought to employ sentences identified from the formative study to test the performance of the model, we found that a significant amount of the sentences that indicated user intents were related to a variety of other applications and functionality used in videoconferences such as presentation, design, internal resource management applications. The CrossTalk system as a prototype did not support these specific features and could not be evaluated using those sentences.

Instead, we opted to construct the dataset using an elicitation survey in which survey respondents (9 students and employees at a large public university) were asked to provide natural expressions for each of the actions that CrossTalk supports. To collect organic expressions, we asked the respondents to recall the expressions they had used to indicate the actions in previous meetings, or if they had not done so before, they were asked to recall expressions that they had heard from other attendees. Respondents were asked to provide as many natural expressions as possible. In total, 350 expressions were collected. For each of the 21 actions CrossTalk supports, 10 expressions were randomly selected, resulting in 210 expressions in total. We selected 3 of the 10 expressions for each action as the training expressions. The training data set also included the description for each action, which consists of 2 sentences, resulting in a training set of 105 sentences. For the remaining 147 collected expressions, we mixed in equal numbers of noisy sentences that were randomly drawn from the Kaggle Human Conversation Training Dataset \cite{ref41HumanConversationtrainingdata}, resulting in a final testing dataset of 294 sentences (provided in supplemental materials). We then employed a weighted KNN (k=5) to recognize the most relevant actions for a testing sentence, weighted by similarity measurement returned by the BERT-based model.

\subsection{Results}
The action prediction model achieved an average precision of 0.89 and an average recall of 0.90, with an average F1-score of 0.89. The evaluation results showed the performance of the current action prediction model used by CrossTalk (RQ2). We found that most error cases resulted from the use of similar expressions that had different meanings under different contexts, e.g., \textit{``I have to go''} is a typical expression one would use to indicate they needed to leave a meeting. However, a similar expression \textit{``I will go first''}, is frequently used in update meetings to indicate that one would like to present first and share their screen. CrossTalk failed to tell them apart. These error cases could be addressed if the semantic similarity calculation considered a longer context of the conversation. Detailed confusion matrix can be found in the supplemental materials. It is also important to note that the reported results do not reflect real-world performance, where speech can be poorly recognized, and short and incomplete phrases are more frequently used.

\section{User Evaluation}

While we attempted to conduct a deployment study to investigate the effects of the proposed interaction on communication and collaboration dynamics, our pilot tests showed that state-of-the-art speech recognition models still produced significant latency and errors for conversational speech in the wild, which influences user perception and reception of the proposed techniques \cite{ref69VoiceInterfacesInEverydayLife}. Therefore, we opt for a controlled user evaluation that focused on obtaining early qualitative feedback on the understanding and impression of the interaction modality of users and uncovering the challenges that need to be addressed for practical use.

\subsection{Participants}
Thirteen participants (9 male, 4 female, age 21 to 39, all fluent English speakers) with experience using remote meetings were recruited to evaluate CrossTalk. Eight participants were from academic backgrounds (5 undergraduate and 3 graduate students) and five were professionals working at different large technology companies (1 engineer, 1 researcher, 1 startup founder, 1 media producer, and 1 marketing manager). Participants were compensated 30 USD for their participation in the 1-hour remote session.

\subsection{Setup}
CrossTalk was deployed on a Heroku server that participants connected to using the Chrome web browser. Participants were required to have a camera, a microphone, and a computer to participate. The experimenter and each participant first met on Zoom for onboarding and then switched to CrossTalk for the study tasks.

\subsection{Procedure}
In each session, the experimenter first introduced the study and the system and then simulated two meeting scenarios with the participant with tasks that the participant needed to complete to experience the proposed technology. The experimenter then concluded the study with a questionnaire and an interview to gather participant’s feedback.

\subsubsection{Introduction (10 minutes)} The experimenter first asked the participants about their most recent remote meeting experience to learn more about each participant's videoconferencing experience and to warm up the participant by grounding them in past experience. Participants accessed CrossTalk through a browser URL and were then introduced to various features of CrossTalk. The next two phases were conducted on the CrossTalk system, where each phase consisted of a teaching portion and open-ended roleplay.

\subsubsection{Meeting Scenario 1: Discussing Places to Go (15 minutes)} Participants engaged in a free-form conversational simulation where the experimenter played a role as a classmate or colleague looking for recommendations on places to go in the participant’s city. During the scenario, participants interacted with map panels, web search panels, and the transcript panel using both manual interaction and language-oriented interaction.

\subsubsection{Meeting Scenario 2: Information Session (15 minutes)} The experimenter and the participants then simulated an organizational meeting to discuss the next steps of a company. During this scenario, PDF documents were shared, and a presentation was given by the experimenter using screensharing. Participants engaged in discussions with the experimenter about content in the PDF and presentation using free-form conversation, during which content recommendations for PDF panels and recorded screen sharing videos were experienced by the participants. Participants were then given a miniature quiz, created on the fly with CrossTalk, on topics in the presentation, PDF, and novel trivia (e.g., \textit{``In what year was Albert Einstein born?''}) and asked to provide the correct answers by interacting with the polling panel.

\subsubsection{Questionnaire and Exit Interview (20 minutes)} Lastly, participants completed a questionnaire about CrossTalk that included questions about its usefulness and usability, using 5-point Likert scale questions. The experimenter concluded the study with a semistructured interview to collect additional feedback on the prototype, focusing on language-oriented interactions as well as issues and concerns with this interaction modality.

\subsection{Study Results}
All participants were able to complete the tasks in the study with ease and responded positively to CrossTalk and found it was intuitive to understand (7/13 strongly agree, 6/13 agree) and easy to use (8/13 strongly agree, 5/13 agree) when the system correctly recognized their speech. In the following sections, we present the participants' responses to key concepts within CrossTalk.

\subsubsection{Reducing Cognitive and Manual Effort during Videoconferences}
Participants appreciated CrossTalk’s ability to interpret their conversation to recommend relevant actions and content and found that it could effectively support communication (5/13 strongly agree, 7/13 agree, 1/13 neutral) and collaboration (5/13 strongly agree, 7/13 agree, 1/13 neutral). Participants perceived these recommendations as \textit{``shortcuts''} (P1) and \textit{``actionable pop-ups''} (P4) and found that the recommendations enabled them to quickly retrieve relevant information. For example, P3 commented that \textit{``It was really cool that I could say words I remembered from the presentation, and it would know where that specific slide was''}. Similarly, P5 also noted that they \textit{``resonated a lot with the reference to people's presentation, pull that up, and go back to places where some important part was''}.

Participants specifically appreciated that relevant recommendations allowed them to maintain the flow of conversation and communication. For example, P7 noted that \textit{``I really liked that we were talking about one place, and it just popped up and asked if we wanted to find it on the map. It felt magical to suddenly have the thing you're talking about appear and you can focus on it to continue the discussion''}.  Similarly, P8 noted that \textit{``the blue dots [which allow users to quickly navigate to the location of a match] is super useful. Sometimes you don't know where it is in a PDF and you have to Control + F and find it. It's great that it just suggests it.''}

P4 reflected on their experience of hosting collaborative videoconferences and found that CrossTalk could significantly reduce their workload. They commented, \textit{``When you lead a conference, especially when you're not just videoconferencing from a conversational standpoint, there're lots of actions you need to take. The ability to surface actionable pop-ups and tasks right in the tool itself reduce the need to switch to a different task. Normally if we're using Zoom and I have to show someone a document. I have to keep in mind where that information is and find it myself. Today, I liked it does it for me. It's not perfect. It's not able to find the right keyword all the time, but that's OK because it removes half the work.''}

In summary, participants enjoyed CrossTalk’s ability to quickly retrieve the information and actions they had in mind, reducing the cognitive and manual effort required of them during videoconferences. Because CrossTalk engaged in the conversation, participants reported that they could rely on CrossTalk to ensure they did not miss important moments or requests in the meeting, as P13 noted \textit{``It also feels very friendly to people who might be multitasking or looking at other things.''} Participants saw themselves using CrossTalk in their daily workflows, such as \textit{``in the workplace for [product] planning''} (P1), \textit{``code review to help them focus everyone's attention to the same place''} (P3), \textit{``[in a group study for] exam preparation''} (P6), or \textit{``talk with my advisor [using CrossTalk] and have [suggestions] in the paper so we can both look at it''} (P9).

\subsubsection{Facilitating Casual and Exploratory Search}
We also observed the reactions of participants to unexpected information recommendations made by CrossTalk. For example, when discussing places to go, participants mentioned types of locations they would like to visit without expressing an explicit intent. CrossTalk, however, still retrieved relevant information and recommended it to the participants. Participants responded positively to these movements. For example, P12 commented that \textit{``It's fun to see what it recommends, even if it's not always right''}. In addition, participants enjoyed that CrossTalk can help them quickly explore a large amount of information \textit{``I really liked today's concept because you could pull up a lot of things at the same time''} (P8). Participants reported that they saw themselves using CrossTalk for casual use cases as \textit{``a trip planning tool''} (P5), for \textit{``tabletop games''} (P12), for a \textit{``book club or movie watching club where you might want to discuss many aspects of the content''} (P1). P13 mentioned this directly: \textit{``I can see myself using this for more casual things with friends to talk about stuff''}. 

\subsubsection{Privacy, Novelty Effect, and Tolerance}
We were also interested in exploring whether CrossTalk can cause privacy concerns as it uses user conversations to provide recommendations. Participants responded positively to the statement that \textit{``I understood that the system offered intelligent recommendations by interpreting conversation''} (7/13 strongly agree, 6/13 agree), and the statement that \textit{``I'm comfortable with the system interpreting my speech to make recommendations''} (6/13 strongly agree, 6/13agree, and 1/13 neutral).

This openness to the system’s transcribing and interpretation of their speech could be for multiple reasons. First, as reported above, CrossTalk demonstrated its value for tasks that were tedious with existing tools. In addition, we found that CrossTalk had a novelty effect on the participants. P7 found it \textit{``magical''} when CrossTalk could provide the right information when they needed it. Similarly, P8 noted that intelligent recommendations were \textit{``really surprising and unexpected''} when inferring their intent. The value CrossTalk can provide, and the novelty effect, can make participants more willing to accept the potential loss of privacy. 

Several other factors could also contribute to the openness of participants, such as sufficient awareness of how their speech was transcribed and used, tolerance to research prototypes than commercial products, or the casualness and informality of the simulated discussion scenarios. For example, P5 mentioned that \textit{``I'm not sure if that would be a privacy issue if you're talking about a person or something controversial''}. 

P12, who rated neutral for being comfortable with the system interpreting their speech, raised the concern that \textit{``I'd want to have the option to remove my data... Anything that's permanent would give me some anxiety. I think people don't always want to be completely on the record. How can you decide when you have control over it?''} This suggests that a future deployment should give the user the ability to control whether their speech is being used for recommendation. The change in the usage ratio of intelligent recommendations in a deployment study can provide valuable information on the adoption of the proposed concepts by users. Nevertheless, these potential factors warrant future research in this direction.

\subsubsection{Latency and Errors with Speech Recognition and Discoverability}

Participants encountered several issues with regard to the usability of natural language interfaces. For example, P6 felt that speech recognition was \textit{``laggy and unresponsive''} which led to an unpleasant experience. This was expected, as state-of-art speech recognition has a considerable amount of latency, making it less suitable for interactive use. Errors in speech recognition can also lead to intent recognition errors that can result in a failure to recommend relevant content and actions.

P2 and P9 both commented on the discoverability of the user interface. Because CrossTalk emphasizes language-oriented interaction with the content, its interface does not contain as many graphical controls as one typically sees in other applications. For example, P4 \textit{``would like to get more information about what to click on and what to use like more traditional systems.''} Similarly, P2 noted that \textit{``off the top of my mind I wouldn't know how to use it, [and I would like] something that tells to click here and there.''} However, P2 found that \textit{``after explanation, it [language-oriented interaction] was more intuitive''}. Such discoverability issues were also found in early research on multitouch interfaces, where gestures could not be discovered by users due to the lack of visual representations depicting their use \cite{ref32PenTouch,ref92ObjectOrientedDrawing}. This problem can be mitigated with dedicated learning systems or through prolonged use over time.

\subsubsection{Summary}
Results from the user study showed that CrossTalk was received positively by participants. The shared information space and the language-based intelligent support offered versatility and flexibility to interact with shared information while reducing the cognitive and manual effort needed during communication and collaboration. The results also uncovered participant concerns and several novel opportunities to improve user experience with natural language interfaces in conversational settings.

Our study has several limitations. As a controlled and small-scale study with simulated meeting scenarios, the findings from our preliminary user study do not provide insights as to how these proposed techniques would be perceived and received in the wild. While participants saw themselves using CrossTalk in a variety of scenarios, meeting dynamics are from more complex and intricate, and can be affected by factors such as settings, number of attendees, relationships among attendees and others. We see CrossTalk as an initial step and a system that the community can leverage to explore these research questions.

\section{LIMITATIONS AND DISCUSSION}

We reflect on the design and evaluation results of CrossTalk and propose several directions that can further the exploration of leveraging natural language to enhance human communication. 

\subsection{Errors of Speech and Intent Recognition}
As found in the study, participants’ experiences with CrossTalk were negatively affected by speech recognition errors, which led to failures in intent detection. To compensate for these errors, CrossTalk supports editing, refining, and recovering through manual interaction as recommended in literature \cite{ref1Siri}. However, discovering, interpreting, and correcting errors can also lead to interruptions in the flow of conversation and unpleasant user experiences. It is important to recognize that natural human conversation is prone to speech recognition errors because of the frequent use of short, incomplete sentences and the overlapping utterances of multiple speakers. However, this limitation could be alleviated with continued improvements in speech recognition. In a deployment study to examine long-term behavior change, one can strategically select the actions that can be most reliably recognized to avoid falling into the trap of speech recognition errors.

\subsection{Latency of Natural Language Interfaces}
While converting speech to text and then extracting commands and intentions from text is a common approach when developing voice input systems \cite{ref46VocalShortcuts,ref51Pixeltone}, there are several layers of latency in speech recognition, natural language processing, and intention detection. Although latency can be mitigated with more performant recognition and understanding of natural languages, such an approach may never be perceived as responsive as people can often anticipate each other’s intentions based on shared knowledge, multimodal cues, and social norms \cite{ref52TurnTaking}. Therefore, the immediate next steps are, first, to identify what latency is acceptable for responsive interactive conversation-based systems, and second, to develop an intention prediction rather than intention detection system based on the context of the conversation. Given the increasingly popular and powerful large language models, it would be interesting to explore whether they can predict users’ unspoken words to significantly reduce latency in existing natural language interfaces.

\subsection{Mapping Natural Language Expressions with Interface Actions at Scale}
Prior natural language interfaces have mostly employed a rule-based approach to map users’ natural language expressions to system functionality \cite{ref46VocalShortcuts,ref51Pixeltone}. While this has enabled the rapid prototyping of natural language interfaces, these limited vocabularies constrain the expressions that can be used to interact with systems. CrossTalk leverages semantic similarity between users’ natural expressions and the descriptions of interface actions, as well as a small set of collected expressions, and demonstrates sufficient performance for action recognition. As mentioned in the formative study, participants tended to think-aloud in a group setting to inform others about the actions they were about to take. This behavior thus presents an opportunity to establish the mappings between natural language expressions to interface actions at scale, by collecting the natural language expressions articulated immediately before or after an interface action. We intend to collect and open-source a dataset of mappings between natural language expressions and interface actions. 

\subsection{Broader Design Space of Intelligent Communication and Collaboration}
While we explored the use of conversational speech in the domain of video-based communication and collaboration, the interaction techniques that were designed can be extended to other communication media such as large interactive displays or augmented and virtual reality for a variety of settings. When using these media, user intentions can be inferred from other signals such as hand gestures, body postures, eye movements, facial expressions, etc. For example, if a user seeks to retrieve a 3D model in virtual reality during a group discussion, their body, head, and eye movements could be used to predict and retrieve the target objects and avoid disrupting the group. Future work can explore the necessary adaption for, as well as the unique challenges and opportunities afforded by, the diverse range of technical and social settings.

\section{Conclusion}
With all of the advances in capacity, speed, and connectivity for transmitting information, there is little contextual sensitivity and intelligence about understanding users’ intents while communicating and collaborating. We explore the use of conversational speech during videoconferences to provide intelligent and context-aware assistance. Based on the findings of a formative study, we proposed a panel substrate that provided an appropriate substrate representation of information to meeting attendees, an intent recognition pipeline that can recognize user intents, and a set of language-oriented interaction techniques that operated on these substrates and intents. The prototype, CrossTalk, implemented these design ideas and showed considerable promise in fostering fluid communication and collaboration. We look forward to expanding the interactions we have explored to a variety of other communication and collaboration settings, as well as fostering collaboration between CSCW, HCI, and AI/NLP researchers to bring this new interaction opportunity to life. 

%

\balance
\bibliographystyle{ACM-Reference-Format}
\bibliography{references}

\appendix
\onecolumn

\section{Formative Study Materials}
\label{app:formative}
\begin{table*}[h]
\centering
    \caption{Recorded meetings there were analyzed. (The 25-hour video category contains meetings of a variety of facilitation and Q\&A styles as a different team led each meeting.)}
    \begin{tabular}{ccccc}
        \toprule
        Setting & Meeting Duration & Attendee & Meeting Activity & Total Duration \\ \midrule
        Industry & 0.5 $\sim$ 1h        & 10 $\sim$ 20  & Application Feature Testing and Critique & 2h  \\ \hline
        Industry & 1h              & 5 $\sim$ 15   & System Design and Code Review & 7h  \\ \hline
        Govt.    & 1 $\sim$ 2h          & 20 $\sim$ 30  & Bid Proposal Evaluation and Discussion & 6h  \\ \hline
        Academic & 1h              & 10 $\sim$ 15  & Team Research Update and Discussion & 25h  \\ \hline
        Academic & 0.5h            & 4 $\sim$ 5    & Discussion and Feedback, Work Allocation & 10h  \\ \hline
        Academic & 0.5 $\sim$ 1h        & 2 $\sim$ 3    & Brainstorm, Work Allocation & 10h  \\ \bottomrule
  \end{tabular}
  \label{table:DemographicsTable}
\end{table*}

\end{document}